\documentclass[twoside]{article}
\usepackage{fleqn,espcrc2}
\usepackage{graphicx}
\usepackage{cite}
\usepackage{times}
\usepackage{amsmath}

\newenvironment{smitemize}{\begin{list}{\labelitemi}{\leftmargin1.5em \itemsep0ex}}{\end{list}}
\newcounter{smenum}
\newenvironment{smenumerate}{\begin{list}{\arabic{smenum}.}{\usecounter{smenum}\leftmargin1.5em \itemsep0ex}}{\end{list}}
\newcommand{\smcaption}{\vspace*{-1cm}\caption}
\renewcommand{\d}{\delta}
\renewcommand{\b}{\beta}

\newcommand{\ra}{\rightarrow}
\newcommand{\rf}[1]{(\ref{#1})}

\title{Center vortices in MCG -- finite-size and gauge-copy effects}

\author{Roman Bertle\address{Inst. f\"ur Kernphysik, 
        Technische Universit\"at Wien,
        Wiedner Hauptstr.~8-10/142, A--1040 Vienna, Austria}%
        \thanks{Talk presented by R. Bertle.
		        Supported in part by FWF 11387-PHY.},
        Manfried Faber\addressmark,
		Jeff Greensite\address{The Niels Bohr Institute,
        Blegdamsvej 17, DK--2100 Copenhagen \O, Denmark},
        {\v S}tefan Olejn\'{\i}k\address{Institute of Physics, Slovak Academy 
        of Sciences, SK--842 28 Bratislava, Slovakia}}

\begin{document}

\begin{abstract}
In SU(2) lattice gauge theory in maximal center gauge, we investigate the dependence of center-projected Creutz ratios and the vortex density on lattice size and the number of gauge copies.
The dependence on the number of copies is rather strong on small lattices, but almost disappears on lattices sufficiently large compared to the expected average vortex thickness.
The center-projected string tension, evaluated on sufficiently large lattices, is in good agreement with the full asymptotic string tension, and the vortex density scales according to the second-order asymptotic-freedom formula.
\vspace{1pc}
\end{abstract}

\maketitle

\section{INTRODUCTION}

Center projection in the direct version of maximal center gau\-ge (DMCG) \cite{dfggo98} is a tool for detecting center vortices in SU($N$) gauge theory.
A number of numerical studies indicate that center vortices are ubiquitous in the QCD vacuum and give rise to the linear confining potential. 
Recently, however, the validity of the procedure based on center projection in MCG has been put in doubt by Bornyakov, Komarov, Polikarpov, and Veselov (BKPV) because of the Gribov copy problem \cite{bkpv00}.
Thus we carefully investigate the dependence of DMCG results on lattice sizes, gauge copies and convergence criteria in SU(2) gauge theory.

In DMCG, the procedure is to maximize
\begin{equation}
       R = \sum_\mu \sum_x \mathrm{Tr}_A[U_\mu(x)]
\label{tomaximize}
\end{equation}   
by an iterative over-relaxation procedure, where $\mbox{Tr}_A[U]$ is the trace of $U$ in the adjoint representation.
Let $R_n$ denote the value of $R$ after $n$ over-relaxation sweeps.
When $R_n$ is judged to have converged according to a criterion of the form
\begin{equation}
       \frac{R_n - R_{n-50}}{R_n} < \d\;,
\label{delta}
\end{equation}
then the link variables $U_\mu(x)$ are projected onto the nearest center elements $Z$ using $Z_\mu(x) = \mathrm{signTr}[U_\mu(x)]$.

When this gauge-fixing procedure is applied to different gauge
copies of a given lattice configuration, it converges not to the unique global maximum of $R$, but to different, local maxima corresponding to different (Gribov) copies.
One way to minimize the gauge copy dependence is to
carry out the over-relaxation procedure on a number $N_{copy}$ of
random gauge copies and perform center projection on the copy with the
largest value of $R$.

Center-projected Wilson loops, Creutz ratios, etc.\ are observables computed from the center-projected link variables.
It was shown in a number of studies \cite{dfggo98}, that 
\begin{smitemize}
\item thin vortex excitations of the projected lattice (``P-vortices'') are located roughly in the middle of thick center vortices on the unprojected lattice;
\item projected Creutz ratios $\chi_{cp}(I,I)$ are close to the asymptotic
string tension on the unprojected lattice (``center dominance'');
\item the density of P-vortices, for $\b \ge 2.3$, scales according
to asymptotic freedom;
\item removing center vortices (located via the projected lattice)
from unprojected lattices also removes confinement and chiral symmetry
breaking, and brings the topological charge to zero \cite{fe99}. 
\end{smitemize}

The question -- why should this procedure work at all? -- was addressed
in ref.\ \cite{fgoy99}.
There it was shown that in the absence of Gribov copies (i.e.\ if the gauge can be fixed to a global maximum of $R$) center projection in maximal center gauge will always locate a thin vortex inserted anywhere on the lattice.
This was dubbed the ``vortex-finding property'' of MCG.
However, the vortices in the QCD vacuum are not thin, but of finite thickness in physical units, and maximal center gauge is plagued with Gribov copies.

Thus we will study in this article
the sensitivity of center-projected Creutz ratios and the P-vortex density
with respect to: (i) the number $N_{copy}$ of used random gauge copies;
(ii) the lattice size; and (iii) the convergence parameter $\d$ in eq.\ \rf{delta}.
We will here only discuss direct maximal center gauge, although there exist alternatives like Laplacian center gauge \cite{aef00} (which is free of the Gribov copy problem).

\section{$\boldsymbol{N_{copy}}$ AND LATTICE SIZE DEPENDENCE}

\begin{figure}
\centering
\includegraphics[width=\linewidth]{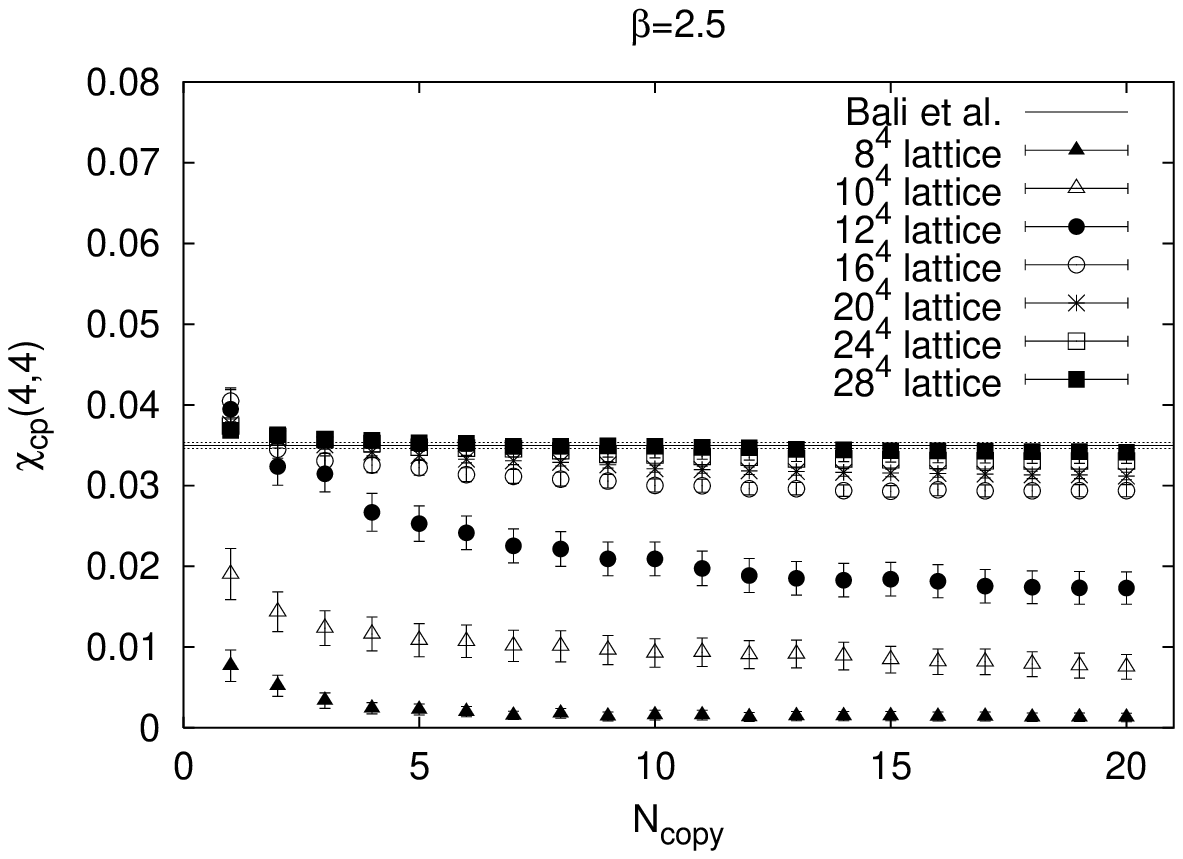}
\smcaption{$\chi_{cp}^{N_{copy}}(4,4)$ vs.\ $N_{copy}$ for $\b=2.5$.}
\label{fig1}
\end{figure}

In the original simulations of ref.\ \cite{dfggo98}, only
three gauge copies were used. Projected Creutz ratios $\chi_{cp}(I,I)$ were
found to be close to the asymptotic string tensions reported by Bali
et al.\ \cite{bss95}, for all $I\ge 2$. In \cite{bkpv00}, however, 
BKPV calculate Creutz ratios in the range $N_{copy} \in [1,20]$, and extrapolate to $N_{copy} \ra \infty$ by fitting their data to the functional form
  $ \chi^{N_{copy}}_{cp}(I,I) = \chi_{cp}(I,I) + 
       {c(I,I)}/{\sqrt{N_{copy}} } $. 
They report that projected string tensions, at
$\b=2.4, 2.5$, underestimate the full string tension by about $20\%$
at $N_{copy}=20$, and by  $30\%$ in the extrapolation to
$N_{copy}\ra \infty$.

   However, BKPV used lattice sizes $12^4$ at $\b=2.3, 2.4$, and $16^4$ at $\b=2.5$, whereas the data reported in ref.\ \cite{dfggo98} was obtained with lattice sizes $16^4$ at $\b=2.3,2.4$, and with $22^4$ at $\b=2.5$.
Therefore the BKPV results may be seriously contaminated by finite-size effects.
To find out, we have repeated the
center-projection calculation at $\b=2.3$ and $\b=2.5$ on a variety of
lattice sizes, for $N_{copy}\in [1,20]$.  For the convergence parameter
in eq.\ \rf{delta}, we have used $\d=2\times 10^{-7}$.

\begin{figure}
\centering
\includegraphics[width=\linewidth]{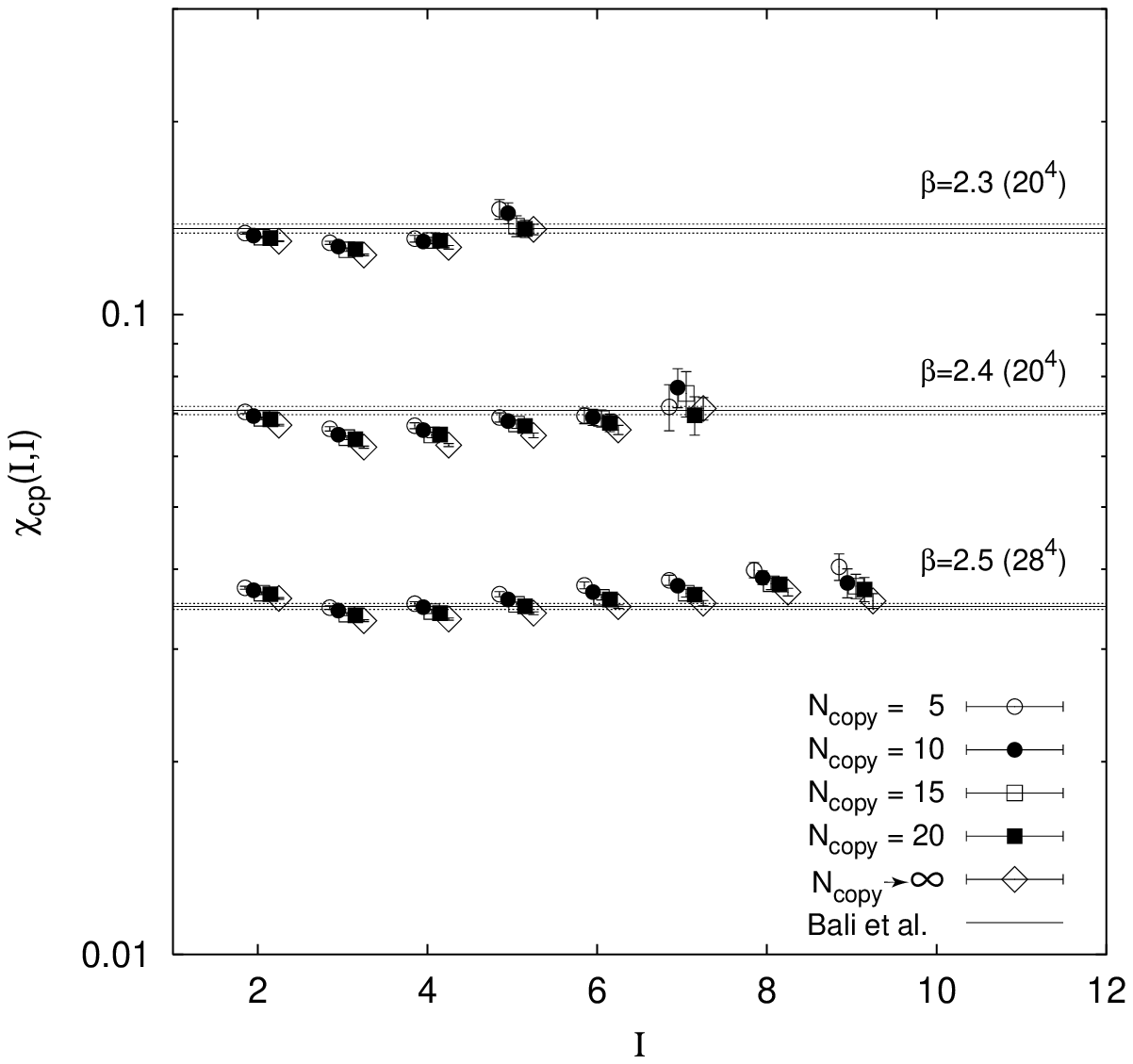}
\smcaption{$\chi_{cp}(I,I)$ for various $N_{copy}$, $\b$ and large $L$.}
\label{creutz}
\end{figure}    

   In Fig.\ \ref{fig1} we show results for the Creutz ratio
$\chi_{cp}^{N_{copy}}(4,4)$ vs.\ $N_{copy}$ at $\b=2.5$, 
for lattice sizes ranging from $8^4$ to $28^4$.
As noted by BKPV, there is a slow downward trend in $\chi_{cp}$ as $N_{copy}$ increases, but this
effect is much more pronounced on smaller lattices than on larger lattices.
And although Creutz ratios on the smaller lattices grossly underestimate
the full string tension $\sigma$ reported by Bali et al.\ \cite{bss95}, the data seems to approach $\sigma$ as the lattice size increases.
These trends in the data are not unique to $\chi_{cp}(4,4)$ at $\beta=2.5$, but are typical of all of our results.

\begin{figure}
\centering
\includegraphics[width=1.0\linewidth]{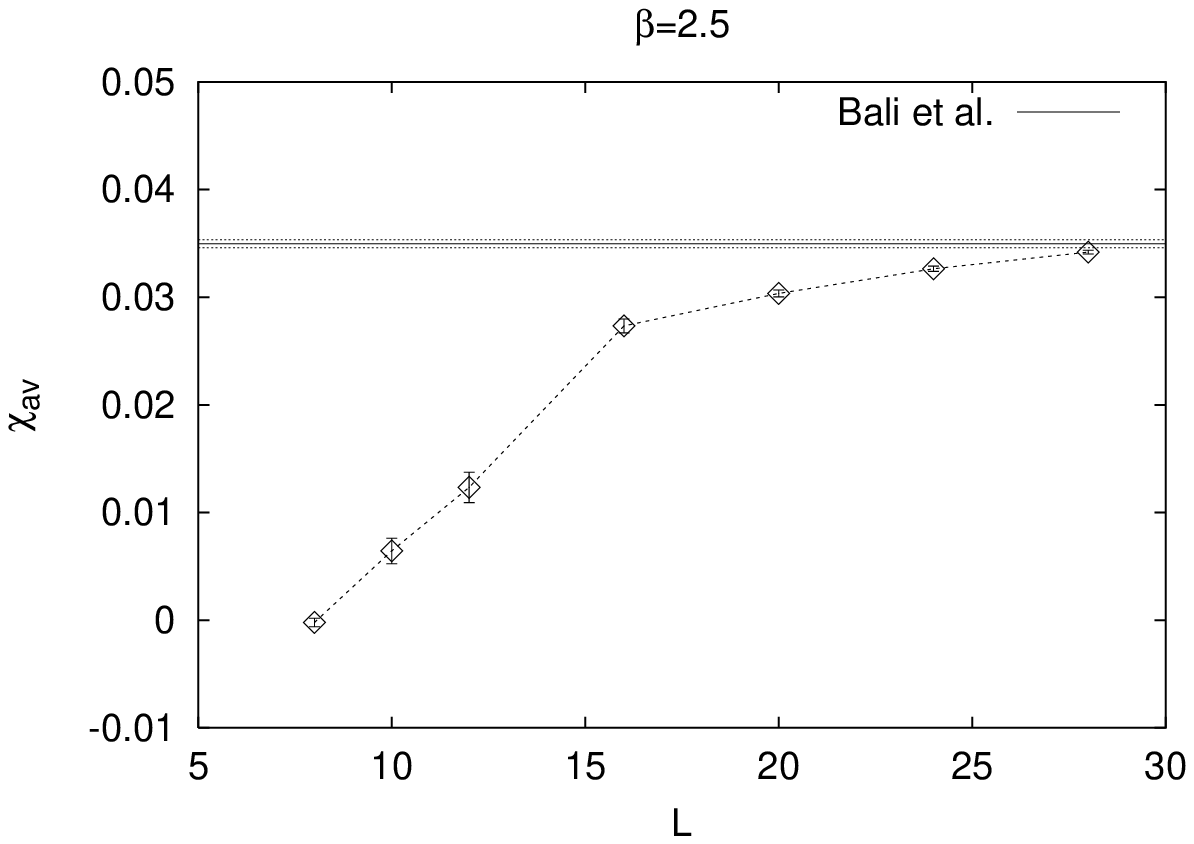}
\smcaption{$\bar{\chi}_{cp}(I,I)$, $I=2-5$, for $N_{copy}\ra \infty$.} 
\label{chi_av}
\end{figure}

  In Fig.\ \ref{creutz} we display the projected Creutz ratios
$\chi_{cp}^{N_{copy}}(I,I)$ for $N_{copy}=5,10,15,20$, and for the extrapolation $N_{copy} \ra \infty$ on the largest lattices we have used:
$20^4$ at $\b=2.3, 2.4$, and $28^4$ at $\b=2.5$.
As usual in MCG, all the $\chi_{cp}(I,I)$ for $I\ge 2$ are close to $\sigma$, and these results are not far from
our earlier results reported in ref.\ \cite{dfggo98}.
In Fig.\ \ref{chi_av}, we show the average $\bar{\chi}_{cp}$ of the projected
($N_{copy}\ra \infty$) Creutz ratios $\chi_{cp}(I,I)$ in the range $I=2-5$.
Note the approach of $\bar{\chi}_{cp}$ to $\sigma$ as lattice size increases.

   In addition we checked that the numerical results are stable
with respect to the constant $\d$ in eq.\ \rf{delta}.
We calculated $\chi_{cp}$ with convergence criteria
$\d=10^{-2},10^{-3},10^{-4},2\times 10^{-7}$ at $\b=2.5$ ($24^4$ lattice).
For the weakest criterion $\d=10^{-2}$, Creutz ratios come out too high, but the results for the two smallest values of $\d$ are fairly consistent,
indicating that these numbers are not far from the $\d \ra 0$ limit.

\section{VORTEX DENSITY AND THICKNESS}

\begin{figure}
\centering
\includegraphics[width=\linewidth]{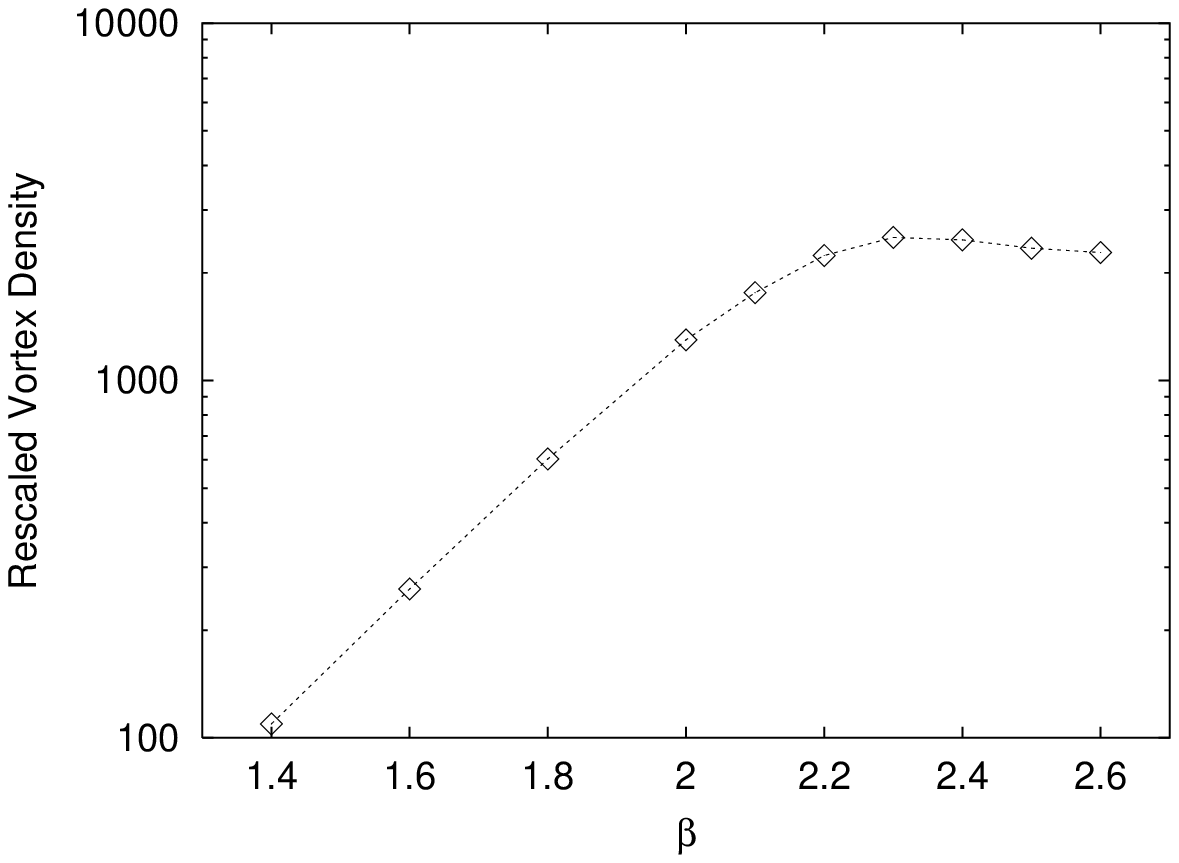}
\smcaption{Rescaled vortex density $\tilde{p}=p/F(\b )$.}
\label{rho}
\end{figure}

The lattice P-vortex density $p$ is the number of P-vortex plaquettes (i.e.\ plaquettes
on the projected lattice with $ZZZZ = -1$), 
divided by the total number of plaquettes on the lattice.
$p$ is proportional to the average area taken up by P-vortices 
per unit lattice volume, and is determined from the center-projected plaquette
expectation value via $p = 1/2 (1 - W_{cp}[1,1])$.
If this quantity scales as predicted by asymptotic freedom, the expression
\begin{equation}
 \tilde{p} \equiv  p \left[\left(\frac{6\pi^2}{11}\b\right)^{102/121} \exp\left(-\frac{6\pi^2}{11}\b\right)\right]^{-1}
\label{F}
\end{equation}
should be constant in the large $\b$ limit.
As can be seen from Fig.\ \ref{rho}, there appears to be good evidence
for this kind of scaling for $\b \ge 2.2$.
The vortex densities at $\b=2.3,2.4,2.5$ 
are taken from the largest used lattices and extrapolated to the $N_{copy}\ra \infty$ limit.
The other values (with $N_{copy}=3$) are taken from our previous work.
It is interesting that the scaling
of $p$, in the range $\b = 2.3-2.5$, is even
better than the scaling of the full asymptotic string tension
$\sigma$ in this range.

In order to avoid large finite size effects, relatively large lattices are required for center projection.
We think that the relevant length scale is associated with the vortex thickness.
Center vortices are
surface-like objects in $D=4$ dimensions which have a finite
thickness in physical units.  The thick vortex surface (or ``core'')
bounds a Dirac 3-volume, which represents the region of discontinuity
of a singular gauge transformation associated with the vortex.  In
ref.\ \cite{fgoy99} we have explained the vortex-finding property of
MCG in terms of the global properties of this
singular gauge transformation.  On a small lattice, with an
extension comparable to the vortex thickness, these global aspects of
the vortex field may be almost absent, and the minimal Dirac 3-volume
could be quite small.  In that case, the argument of ref.\ \cite{fgoy99}
breaks down. Thus we expect center projection to be less
effective at finding vortices on small lattices, leading to
underestimates of both $\chi_{cp}$ and $p$.

  If there is some truth in this explanation, center
projection works well only for lattices which are large compared
to the vortex thickness.
This thickness can be deduced from
\begin{smenumerate}
\item the ratio of ``vortex-limited'' Wilson loops;
\item the vortex free energy as a function of lattice size;
\item the adjoint string-breaking length.
\end{smenumerate}
Vortex-limited Wilson loops are defined in the following way \cite{dfggo98}: $W_n(C)$
is a Wilson loop evaluated on a sub-ensemble of unprojected
configurations, selected 
so that precisely $n$ P-vortices, in the corresponding center-projected
configurations, pierce the minimal area of the loop.
For $W_1(C)$ we can further require that the negative P-vortex plaquette
lies at (or touches) the center point of the rectangular loop.  It is
then expected that 
     $ W_1(C) / W_0(C) \ra -1 $
in the limit where the vortex core is entirely contained within
the loop.
At $\b=2.3$ the vortex appears to almost fit inside a $5\times 5$ loop \cite{dfggo98}, which leads to a rough estimate of the vortex radius of about 3 lattice spacings.
A diameter of 6 lattice spacings at $\b=2.3$ corresponds to a vortex thickness of about one fermi.

   A second estimate is obtained from the recent numerical calculation of
vortex free energy vs.\ lattice size \cite{kt00c}. The vortex free energy is 
close to zero when the lattice extension is greater than the vortex
thickness, and this again gives an estimate for the vortex thickness of
a little over one fermi. Finally, if confinement is due to center
vortices, then an $R\times T$ Wilson loop in the
adjoint representation must change from a (Casimir scaling) area-law
falloff to a (color-screening) perimeter-law falloff for $R$ greater than the vortex thickness \cite{fgo98}.
The adjoint string-breaking distance has been measured to be $1.25$ fm\cite{fp00}, which is roughly consistent with the other two estimates.

   At $\b=2.5$, one fermi corresponds to 12 lattice spacings.
BKPV used at $\b=2.5$ only a $16^4$ lattice which may simply be not large enough compared with the vortex thickness in order to reliably identify vortices using center projection.

The average distance between vortices is given by the vortex density.
The P-vortex density, discussed above, is in fact an overestimate of the
actual center vortex density because P-vortices fluctuate within the
thick vortex core \cite{bfgo99}.
A more accurate estimate of the center vortex
density is arrived at by either ``smoothing'' the P-vortex surfaces, or else directly from the string tension using $f=\frac{1}{2} (1 - e^{-\sigma})$.
The two estimates agree fairly well \cite{bfgo99}, and for $\b=2.3$ we
find $f=0.063$.  This implies an average distance of $f^{-1/2}\approx
4$ lattice spacings between the centers of vortex cores piercing a
plane.  Since we have already estimated the vortex thickness at
$\b=2.3$ to be about 6 lattice spacings, there must be
some overlap between vortex cores.
There is nothing in principle wrong with that; vortex cores
are not impenetrable objects, and their long-range effects are
associated with Dirac
3-volumes, rather than the detailed structure
of the core.
These findings indicate that the QCD vacuum is more like a liquid of vortices than a dilute gas.

\section{CONCLUSIONS}

   We have found that center-projected lattices are more sensitive to
finite size effects than unprojected lattices.
Precision results for center-projected Creutz ratios $\chi_{cp}$ require lattice sizes which are large compared to the physical vortex thickness of $\sim 1 \mathrm{fm}$.
The $N_{copy}$-dependence for $\chi_{cp}$ reported by BKPV \cite{bkpv00} is greatly reduced as lattice size increases.
On the largest lattices we have used, and with the extrapolation to the $N_{copy}\ra \infty$ limit, our results lie quite close to the asymptotic string tension, and are stable with respect to the gauge-fixing convergence criterion.
For the vortex density, we find good evidence for asymptotic scaling.\footnote%
{After the conference, a new paper of Bornyakov et al.~\cite{bkp00}
appeared. Using the 
method of simulated annealing, they find even higher 
maxima of $R$, eq.\ \rf{tomaximize}, and the center-projected string tension
smaller than the physical one. It seems clear that a modification of
the gauge-fixing procedure and/or condition is required.}

\bibliographystyle{kphunsrt}
\bibliography{bertle}

\end{document}